\def\rect#1#2{{\vcenter{\vbox{\hrule height.3pt
	    \hbox{\vrule width.3pt height#2truecm \kern#1truecm
	    \vrule width.3pt}
	    \hrule height.3pt}}}}

\def\aversim#1#2{\lower3pt\vbox{\baselineskip0pt \lineskip-.1pt
    \ialign{$\mathsurround=0pt #1\hfil##\hfil$\crcr#2\crcr\sim\crcr}}}

\documentstyle[12pt]{article}
\begin{document}
\begin{center}
\LARGE
Folding transitions of the triangular lattice with defects
 ~\\
 ~\\
 ~\\
 ~\\
\normalsize
 ~\\
 ~\\
\normalsize
Emilio N.M. Cirillo {\it and} Giuseppe Gonnella\\
{\it Dipartimento di Fisica dell'Universit\`{a} di Bari} and\\
{\it Istituto Nazionale di Fisica Nucleare, Sezione di Bari\\
via Amendola 173, 70126 Bari, Italy}
 ~\\
 ~\\
\normalsize
Alessandro Pelizzola\\
{\it Dipartimento di Fisica del Politecnico di Torino} and \\
{\it Istituto Nazionale per la Fisica della Materia \\
c. Duca degli Abruzzi 24, 10129 Torino, Italy}
 ~\\
 ~\\
\end{center}
\vskip 2.5cm
\begin{abstract}
A recently introduced model describing the folding of the triangular
lattice is generalized allowing for defects in the lattice and written as
an Ising model with nearest--neighbor and plaquette interactions on the
honeycomb lattice. Its phase diagram is determined in the hexagon
approximation of the cluster variation method and the crossover from the
pure Ising to the pure folding model is investigated, obtaining a quite rich
structure with several multicritical points. Our results are in very good
agreement with the available exact ones and extend a previous transfer
matrix study.
\end{abstract}
\newpage

\addtolength{\baselineskip}{\baselineskip}

\section{Introduction}

Polymerized membranes are two-dimensional generalizations of linear
polymers; the large variety of
 possible applications and the extension of one-dimensional
 statistical properties to $2D$ objects justifies  recent numerous
studies on these systems \cite {NPW,KN,NP,PKN,DG,AL,GM,BEW}.
Differently from linear polymers, polymerized membranes are expected to show
different long-distance behaviours depending on the microscopic
characteristics of the system.
In particular, rigid polymerized membranes are expected to be stable in a flat
phase \cite {NP} which has no analogue in polymer systems.

Models of polymerized membranes consist of
networks with fixed connectivity fluctuating in an embedding space.
The length of the bonds can vary  with an energy cost diverging
for increasing lengths, while
the rigidity of the network is described by
a bending energy term  favouring flat configurations.
When  excluded volume effects are not considered, as it will be in
the model of  this paper,  it has been shown \cite {KN,NP,PKN,DG,AL,BEW}
that by varying
the strength $K$ of the bending  term,  a critical
transition arises separating a  flat phase at large $K$ from
a crumpled
phase  at small $K$.

The case of a triangular network  embedded in a $d$-dimensional space
where the bonds are of fixed  length has been first considered in
\cite{KJ}.
Here the only degrees of freedom correspond to the
possible ways of folding  the network. In \cite {KJ} the problem
has been studied in a $2d$ embedding space. In this simplified case
the normal vectors to the  triangles
point either up or down in some direction, suggesting a description
of the system
in terms of Ising spin variables.
The entropy of this folding problem has been recognized
 \cite{epl} to be the same entropy of the problem of colouring with three
colours the  bonds of the hexagonal lattice,
 which has been exactly calculated  in \cite {B}.
In the description of the model in terms of Ising variables a  bending rigidity
can be  easily  defined  \cite {KJ}; in \cite {pre} a first-order
$2d$ folding
transition  between a crumpled and a flat phase has been  found at a finite
value of the bending strength $K$.

In this paper we consider again the $2d$ folding problem of a triangular
network; we apply the cluster variation method (CVM) \cite{nim,an,morita}
to complete the study of \cite {pre}
 analyzing  the phase diagram of the model in the whole plane
$K,h$ where $h$ is
a symmetry breaking field distinguishing
between up and down triangles. The folding problem can be
expressed as a vertex model or
an Ising problem with a local  constraint on the possible
spin configurations \cite {epl}. We have also studied the system by
progressively
relaxing the  constraint until the standard 	Ising model is obtained.
The relaxing of the constraint corresponds to accept folding configurations
with endpoints in the folding lines, that is
 cuts between adjacent
triangles of the network.
 Therefore intermediate models between the Ising model
and the folding model of \cite {KJ,epl,pre}  could describe realistic
polymerized membranes where defects in the connectivity rules are
present.

The cluster variation method is known
to be very accurate in describing the phase diagram of magnetic systems
\cite{cvmgen}.
Also in this case of a  constrained magnetic system, the CVM
gives an accurate  description of the system.
For example,  the  $T=\infty$ entropy calculated with the CVM
is $\ln q$, $q=\sqrt{13}/3\approx 1.2019$
which has
to be compared with the exact value $q=1.2087$ \cite {B}.
In the whole plane $K,h$, in addition to the first order transition
between the flat  and the folded
disordered phase \cite{pre},
we have  found at
negative $K$ a critical
transition between the disordered phase and a folded phase
with staggered antiferromagnetic order.
When  the constraint  is progressively relaxed
 the topology of the
phase diagram is found to evolve in  quite a complex way, resembling
at negative $K$ the phase diagram of the Ising model for metamagnets
\cite {KC}, until
 the  standard Ising model phase diagram is recovered.

The paper is organized as follows. In section 2 the model of folding is defined
in terms of spin variables and the CVM approximation scheme  used for
studying the phase diagram is briefly described.
In  section 3 the phase diagrams of the folding system with
the constraint progressively
relaxed are given. In section 4 our results will be briefly discussed.

\vskip 2 truecm
\section{The model and the method}

We briefly describe the model of folding studied in \cite {KJ,epl,pre}.
In a $2d$ embedding space
 the normal vectors to the triangles of a regular network
can be represented by Ising spins
$s_i = \pm 1$, where $i$ denotes a site of a honeycomb
lattice (the dual of the triangular lattice), in such a way that a $+$
(respectively $-$) spin corresponds to
a normal vector pointing up (down). Following \cite{pre},
we  consider a bending rigidity $K$, which
measures the energy
cost (in units of $k_B T$) of a fold between adjacent triangles, and a
\lq\lq magnetic\rq\rq\ symmetry--breaking field $h$, conjugate to the
normal vectors.
Furthermore, in order to study the crossover
from the Ising model to the pure folding model, we shall allow the spins
around a hexagon to violate the local constraint
\begin{equation}
\sum_{i \in {\rm hexagon}} s_i = 0 \quad {\rm mod} \quad 3
\label{constr}
\end{equation}
(only 0, 3 or 6 minus spins allowed in a hexagon),
which was introduced in \cite{epl}.
When the constraint Eq.\ \ref{constr} is verified, the model can be expressed
as a 11-vertex model \cite {epl}.
In our generalized folding model, a violation of the constraint
Eq.\ \ref{constr} will cost an energy $L>0$, such that $L = 0$ corresponds to
the
ordinary Ising model, while in the limit $L \to \infty$ one recovers the
pure folding model.
We are thus led to consider the following hamiltonian:
\begin{equation}
- \frac{{\cal H}}{k_B T} = K \sum_{\langle i j \rangle} s_i s_j
+ h \sum_i s_i
+ L \sum_{{\rm hexagons}} \delta \left( \{s_i\}_{i \in {\rm hexagon}}  \right),
\label{ham}
\end{equation}
the first sum is over nearest neighbors and $\delta$ is a function
equal to 1 only when the constraint Eq.\ \ref{constr} is satisfied and
zero otherwise;
the hamiltonian above will be studied in the hexagon approximation of the
cluster
variation method.

The CVM, in its modern formulation \cite{an,morita}, is based on the
minimization of a free energy density functional which is obtained
by a truncation of the cluster (cumulant) expansion of the
corresponding functional appearing in the exact variational
formulation of statistical mechanics.
In the hexagon approximation for the honeycomb lattice the largest clusters
appearing in the expansion are the hexagons, and the
approximate free energy density one has to minimize, determined according
to \cite{an}, has the form
\begin{eqnarray}
f(\rho_6) & = & - \frac{3}{2} K {\rm Tr}(s_1 s_2 \rho_2(s_1,s_2))
- h {\rm Tr}(s_1 \rho_1(s_1)) -
\frac{1}{2} L \textstyle{\sum^\prime} \rho_6( \{ s_i \} ) \nonumber \\
&& + \frac{1}{2} {\rm Tr}(\rho_6 \ln \rho_6) -
\frac{3}{2} {\rm Tr}(\rho_2 \ln \rho_2)
+ \frac{1}{2}{\rm Tr}(\rho_{1A} \ln \rho_{1A})
+ \frac{1}{2}{\rm Tr}(\rho_{1B} \ln \rho_{1B}) \nonumber \\
&& + \lambda({\rm Tr} \rho_6 - 1),
\label{free}
\end{eqnarray}
where Tr stands for trace, $\sum^\prime$ denotes summation
over the hexagon configurations that satisfy the constraint Eq.\ \ref{constr},
$\lambda$ is a Lagrange multiplier which ensures the normalization of
$\rho_6$ (normalization of the site and pair density matrices follows by
definition, see below) and
$\rho_{1A(B)} \equiv \rho_{1A(B)}(s_1)$, $\rho_2 \equiv \rho_2(s_1,s_2)$ and
$\rho_6 \equiv \rho_6(s_1,s_2,s_3,s_4,s_5,s_6)$ are the site, pair and
hexagon density matrices respectively.
The spin configuration of each cluster can be used as a label for the
corresponding density matrix, because, as for any classical model, the
density matrices are diagonal.

In writing Eq.\ \ref{free} we have introduced two site density
matrices, $\rho_{1A}$ and $\rho_{1B}$, corresponding to the two
interpenetrating sublattices in which the honeycomb lattice can be divided,
which will be needed to investigate the case of antiferromagnetic
coupling $K < 0$. These density matrices can be defined as partial traces
of the pair matrix, as follows
\begin{eqnarray}
\rho_{1A}(s_A) & = & \sum_{s_B} \rho_2(s_A,s_B) \nonumber \\
\rho_{1B}(s_B) & = & \sum_{s_A} \rho_2(s_A,s_B),
\label{rho1}
\end{eqnarray}
adopting the convention that the first spin in the argument of $\rho_2$
always belongs to sublattice $A$. With this assumption, $\rho_2$ can be
defined as a (symmetrized, for convenience) partial trace of $\rho_6$ by
\begin{eqnarray}
\rho_2(s_A,s_B) & = & \frac{1}{6}
\sum_{s_A^\prime,s_A^{\prime\prime},s_B^\prime,s_B^{\prime\prime}} \left[
\rho_6(s_A,s_B,s_A^\prime,s_B^\prime,s_A^{\prime\prime},s_B^{\prime\prime}) +
\rho_6(s_A^\prime,s_B,s_A,s_B^\prime,s_A^{\prime\prime},s_B^{\prime\prime})
\right. \nonumber \\
&& +
\rho_6(s_A^\prime,s_B^\prime,s_A,s_B,s_A^{\prime\prime},s_B^{\prime\prime})
+ \rho_6(s_A^\prime,s_B^\prime,s_A^{\prime\prime},s_B,s_A,s_B^{\prime\prime})
\nonumber \\
&& \left. +
\rho_6(s_A^\prime,s_B^\prime,s_A^{\prime\prime},s_B^{\prime\prime},s_A,s_B) +
\rho_6(s_A,s_B^\prime,s_A^{\prime\prime},s_B^{\prime\prime},s_A^\prime,s_B)
\right],
\label{rho2}
\end{eqnarray}
where the spins in the argument of $\rho_6$ follow each other
counterclockwise in the hexagon, and the first one is on the $A$ sublattice.
In terms of the site and pair density matrices one can easily define the
uniform and staggered order parameters
\begin{eqnarray}
M & = & \frac{1}{2} \left[ {\rm Tr} (s \rho_{1A}(s)) + {\rm Tr} (s
\rho_{1B}(s)) \right], \nonumber \\
M_S & = & \frac{1}{2} \left[ {\rm Tr} (s \rho_{1A}(s)) - {\rm Tr} (s
\rho_{1B}(s)) \right],
\end{eqnarray}
and the nearest neighbor correlation function $c = {\rm Tr}(s_A s_B
\rho_2(s_A, s_B))$.

With the definitions above our free energy can be regarded as a function of
$\rho_6$ only and taking the derivatives with respect to the generic
element of $\rho_6$ we find, after some algebraic manipulation, the
stationarity conditions
\begin{eqnarray}
\rho_6(s_1,s_2,s_3,s_4,s_5,s_6) & = &
\exp \left[ -\lambda + \frac{K}{2} \sum_{i=1}^6 s_i s_{i+1} +
\frac{h}{3} \sum_{i=1}^6 s_i + L \delta
\left( \{s_i\}_{i \in {\rm hexagon}}  \right) \right] \nonumber \\
&& \times \left[ \rho_2(s_1,s_6) \rho_2(s_1,s_2) \rho_2(s_3,s_2)
\rho_2(s_3,s_4) \rho_2(s_5,s_4) \rho_2(s_5,s_6) \right]^{1/2} \nonumber \\
&& \times \left[ \rho_{1A}(s_1) \rho_{1B}(s_2) \rho_{1A}(s_3) \rho_{1B}(s_4)
\rho_{1A}(s_5) \rho_{1B}(s_6) \right]^{-1/3},
\label{nim}
\end{eqnarray}
where $s_7 \equiv s_1$ and
$\lambda$ has to be determined by solving the normalization condition
${\rm Tr} \rho_6 = 1$. This set of equations, together with the definitions
Eqs.\ \ref{rho1} and \ref{rho2} is known as the natural iteration
method \cite{nim}, because it can be solved by simply iterating the
equations above, and always converges to a
local minimum of the approximate free energy. To find the global minimum it
is therefore enough to start the iteration with different sets of initial
conditions, one for each of the expected phases.

It can be easily recognized that the elements of $\rho_6$ are not all
independent, since when two spin configurations are related by simmetry
(rotation and/or reflection) the corresponding elements are degenerate. In
Tab.\ 1 we have listed the 20 independent (up to normalization)
configurations, together with their multiplicities, for the most general
case, corresponding to the antiferromagnetic phase (see below). In the
ferromagnetic phase we have sublattice invariance, which implies (using the
same symbol for a configuration and its density)
\begin{equation}
z_3 = z_4, \qquad z_5 = z_6, \qquad z_9 = z_{10}, \qquad z_{11} = z_{12},
\qquad z_{15} = z_{16}, \qquad z_{17} = z_{18}, \qquad z_{19} = z_{20}.
\end{equation}
In the disordered phase, and for $h = 0$, we have also invariance under
spin inversion, from which one can derive the additional relations
\begin{equation}
z_1 = z_2, \qquad z_3 = z_5, \qquad z_7 = z_8, \qquad z_9 = z_{11}, \qquad
z_{13} = z_{14}.
\end{equation}
Finally, in the limit $L \to \infty$ (pure folding), the densities
corresponding to configurations which violate the constraint Eq.\ \ref{constr}
vanish, i.e. $z_i = 0, \quad i = 3, 4, \ldots 14$.

\vskip 2 truecm
\section{Results}

In this section we present our results for
the phase diagram  of the model Eq.\ \ref{ham}.

First of all, we consider the case $L = \infty$, $K = h = 0$, studied
in \cite{KJ,epl,B,cic}. For $L = \infty$, $h = 0$ and unbroken spin--flip
symmetry
the only non--vanishing elements of $\rho_6$ are
$z_1 = z_2$, $z_{15} = z_{16}$, $z_{17} = z_{18}$ and $z_{19} = z_{20}$. Then
the stationarity condition Eq.\ \ref{nim} can be solved exactly, yielding
\begin{eqnarray}
z_{19} & = &
\left[2 \left( \alpha^3 + 3 \alpha^2 + 6 \alpha + 1 \right) \right]^{-1},
\nonumber \\
z_{1} & = & \alpha z_{15} = \alpha^2 z_{17} = \alpha^3 z_{19}, \nonumber \\
\alpha & = & \frac{2 - u - \sqrt{3 - u - u^2}}{u - 1}, \qquad u = \exp{(2K)}.
\label{soldis}
\end{eqnarray}
For $K = 0$ this reduces to $8 z_1 = 4 z_{15} = 2 z_{17} = z_{19} = 4/39$,
corresponding to an entropy per site $s = \ln q$ with $q = \sqrt{13}/3
\simeq 1.2019$, whereas the exact result \cite{B} is
$q = \displaystyle{\frac{\sqrt{3}}{2\pi}\Gamma\left(\frac{2}{3}\right)^{3/2}}
\simeq 1.2087$, and to a negative (antiferromagnetic) nearest neighbor
correlation $c = -1/3$, for which no exact result is
available. The very good agreement (within 0.6 \%) of our estimate of the
entropy with the exact value gives us confidence in applying the cluster
variation method to the present model.

Let us now turn to the analysis of the phase diagrams in the plane $K,h$
at different values of $L$. They are symmetric with respect to the axis $h=0$
and it is sufficient to describe them at positive values of $h$.

In Fig. 1 it is shown the phase diagram of the pure folding problem
($L=\infty$).
At sufficiently large values of $h$ the flat phase with $M \equiv 1$ is
always stable. It is remarkable that the CVM always yields $M$ exactly
equal to 1 in this phase, as conjectured in \cite{pre}. This result is
presumably exact, since, as observed in \cite{epl}, the flat (ferromagnetic)
ground state has no local excitations. This is no longer true,
of course, for finite values of $L$.

The two  flat phases with magnetization $M = \pm 1$ coexist
at $h=0$ for $K \ge K_0^{cr}(\infty)=0.1013$, to be compared with
the value $K=0.11 \pm 0.01$ of \cite {pre}. This first order transition
point can be obtained by requiring that the free energy densities of the
ordered and disordered phase take the same value, i.e.
by solving the equation $f_{\rm ORD}(K) = f_{\rm DIS}(K)$. Here
$f_{\rm ORD}(K) = - 3 K/2$ is the (presumably exact) free energy density of
the ordered phase, as in \cite{pre}, obtained by observing that this phase
has vanishing entropy and neglecting the diverging $L$ contribution, which
is common to both phases. On the other hand, by substituting the disordered
phase density matrix elements Eq.\ \ref{soldis}
in the expression for the free energy we obtain
\begin{eqnarray}
f_{\rm DIS}(K) & = & - \frac{3}{2} K \frac{\alpha - u}{\alpha + u} +
\frac{1}{2} \ln z_{19} + 3 z_{19} (\alpha^3 + 2 \alpha^2 + 2 \alpha) \ln
\alpha + \nonumber \\
&& - 3 {\cal L} \left[ \frac{\alpha}{2 (\alpha + u)} \right]
- 3 {\cal L} \left[ \frac{u}{2 (\alpha + u)} \right] - \ln 2,
\end{eqnarray}
where ${\cal L}(x) = x \ln x$.

At smaller values of $K$, the  coexistence line at $h=0$ separates
into two branches where the disordered folded
phase with $M \approx 0$ coexists with the  flat phases. The intersection of
the upper branch with the axis $K=0$ is at $h=0.18495\pm 0.00005$ to be
compared with the value $h=0.189$ found in \cite{pre}. In the disordered
phase $M$ vanishes only at $h = 0$, but this might be a consequence of our
approximation (see also the discussion in the next section).

Then, at  sufficiently negative  value of $K$, there is a
critical transition between the disordered folded phase
and a folded antiferromagnetic phase with staggered order parameter
$M_S \ne 0$ and $M\approx 0$.
This transition is represented by the almost vertical broken line of
Fig. 1, which intersects the horizontal axis at $K=-0.284$.
This completes the phase diagram shown in Fig. 6 of \cite {pre}.

In order to understand how this phase diagram evolves while the constraint
is relaxed we begin by plotting in Fig. 2 (in the $L,K$ plane)
the transition line  $K_0^{cr}(L)$ which separates, at $h=0$,
the disordered folded phase with $M=0$ and the flat phases with $M \ne 0$.
At large values of $L$ the transition is first-order and the curve
tends to the asymptotic value  $K_0^{cr}(\infty)=0.1013$.
At small values of $L$  the transition is critical as in the Ising model (L=0).
At $L=1.359, K=0.3038$
there is a  tricritical point where the transition changes its behaviour.
In our approximation scheme the critical line and the tricritical point are
obtained as the solution of suitable analytical equations \cite{peli}.
The derivation
of these equations
is almost straightforward but quite cumbersome, and we have omitted it.

The topology of the phase diagram in the plane $K,h$ remains the same until the
value $L_{\rm b} = 1.75\pm0.05$  is reached. For smaller values of $L$,
in the range $1.359<L<L_{\rm b}$,
the typical phase diagram  is shown in
Fig. 3, where the particular value $L=1.6$ has been chosen.
The difference with the case $L=\infty $ is that now the first-order
transition lines between the disordered and the flat phase are
interrupted in some range.
The  two critical points
limiting the upper and the lower branch of the interrupted
first-order line
(at positive $h$) have
respectively coordinates
$K=0.175,h=0.076$ and $K=0.262,h=0.0009$.

When $L$ still decreases, the branches at smaller $|h|$
 (see  the inset of Fig. 3)
 become shorter and shorter
until they collapse on the axis $h=0$ at the tricritical point of Fig. 2
(L=1.359). In this situation the phase diagram assumes the topology of
Fig. 4 ($L=1.3$), with the two surviving  branches separating
phases with different magnetizations.

By diminishing  further the value of $L$  the branches of Fig. 4
become less pronounced and at $L=1.1$  they are not distinguishable
anymore. Therefore, on the transition line limiting the antiferromagnetic
phase, two tricritical points (in symmetrical positions with respect to
the $K$ axis) separate the first-order
behaviour at more negative values of $K$ from the critical behaviour
close to the $K$ axis. At $L=1.1$ the coordinates of the tricritical
point at $h>0$ are $K=-0.6198$, $h=1.795$.
These tricritical points move towards  larger values
of $|h|$ when  $L$ decreases. It is not clear from our
calculations if  these points disappear at a
singular value of $L$
or if they move continuously towards $|h|=\infty$ when $L\rightarrow 0$.
At $L=0$ the standard phase diagram
of the Ising model in a magnetic field is recovered
with a critical transition always limiting the antiferromagnetic phase \cite
{KC}.
This phase diagram is shown in Fig. 5. The intersection of the critical
line with the axis $h=0$ is at $K=-0.6214$ which has to be compared with the
exact critical Ising coupling on the hexagonal lattice given by $|K_{\rm c}| =
\displaystyle{\frac{1}{2}} \ln (\sqrt{3} + 2) \simeq 0.6585$ \cite {Hi}. The
point $K=K_0^{cr}(0)=0.6214$ is the usual ferromagnetic Ising critical point.

\vskip 2 truecm
\section{Discussion and conclusions}

In this paper we have studied the phase diagram of a constrained
Ising spin model describing a network of equilateral triangles
embedded in a $2d$ space. The constraint is due to the fact
that during the folding the local connectivity properties
of the network have to be preserved. This study has been
done by applying the cluster variation method to the hamiltonian defined in
Eq.\ \ref{ham}. In Eq.\ \ref{ham} the term proportional to $L$ favours
the spin configurations verifying the constraint which are the
only surviving in the limit $L\rightarrow \infty$.
In order to study networks with defects consisting of endpoints in
the folding lines, we have also studied the model Eq.\ \ref{ham}
at finite values of $L$ by progressively relaxing the constraint
until to consider the case $L=0$ corresponding to the usual Ising model.

Our results can  be summarized in this way.
In the case of the pure folding
problem
we first observe the  good agreement between our evaluation of the
entropy $s = \ln q$ with $q \simeq 1.2019$ at $K=0,h=0$
and the exact result $q = 1.2087$ of \cite {B}. We also predict
for the nearest neighbor correlation at $K=0, h=0$
the value $c=-1/3$.
Results concerning other regions of the phase  diagram
in the plane $K,h$ are also in good agreement with results obtained in
\cite {pre}.
We have completed the results
of \cite {pre} by studying the phase diagram of the model Eq.\ \ref{ham}
also for negative values of the bending rigidity $K$.
A critical transition between the disordered folded phase and a folded
phase with antiferromagnetic order has been found. The critical value of $K$
for this transition at $h=0$ is $K=-0.284$. We conclude the discussion of the
pure
folding case  making some comments about
the hypothesis advanced in
\cite{pre} that $M$ always
equals 1 in the ordered phase and 0 in the disordered phase.
Our results confirm that $M \equiv 1$ in the ordered phase, a
result which is almost certainly exact, but yield $M = 0$ in the disordered
phase only for $h = 0$, while close to the transition with the flat phase
we have $M \sim
10^{-2}$. An order parameter of the same order of magnitude can be obtained
by extrapolating with standard methods (Shanks transform, alternating
$\epsilon$--algorithm and Pad\'e approximants \cite{barber,guttmann})
the results from the transfer matrix method proposed in \cite{pre},
obtained with strip widths in the range 2 to 8. It would be very
interesting to see how the order parameter varies when one considers larger
clusters for the CVM or larger strips in the transfer matrix method, but
this is beyond the purpose of the present paper. In our opinion, however,
this issue cannot be settled numerically in a definitive way, and a
rigorous argument would be welcome.

At finite $L$, the evolution of the phase diagram from $L\rightarrow\infty$
to $L=0$ is shown in Figs. 1,3-5.
It is curious to observe  the analogies between the antiferromagnetic
region of  these phase diagrams
  and the
 phase diagrams appearing in the
study of the
 Ising spin model for metamagnets \cite {KC}.
The hamiltonian of the
Ising spin model for metamagnets includes an antiferromagnetic
coupling $J$ between nearest neighboring sites, a ferromagnetic coupling
$J'$
between next-to-the-nearest neighboring sites and an external magnetic
 field. When the ratio $\epsilon =J/J'$ is in the range $0 <\epsilon < 3/5$,
the phase
diagram of the metamagnet is similar for example to that of Fig. 4, but with
the first-order branches  inside the
antiferromagnetic phase. The branches  terminate at critical points where two
antiferromagnetic phases  with
different net magnetizations cease to coexist \cite {KC}. These
analogies can be understood by observing that the constraint energy
proportional to $L$ includes an effective next-to-the-nearest  neighboring site
interaction in the hexagonal lattice.

In conclusions, we observe that in this paper the CVM has been applied
to a vertex model giving results in excellent
agreement with numerical and exact previously known results.
This is encouraging for applying the CVM to the study of other
vertex models or colouring problems. In particular, in \cite {Bos}
an extension of the model of \cite {KJ,epl,pre} which describes
a possible embedding  of a triangular network in a $3d$ space has been
formulated as a 98-vertex model. We think that the phase diagram
of the model of \cite {Bos} could be efficiently studied
by applying the CVM, and work is in progress along these lines.

\newpage
\thispagestyle{empty}
\Large
\begin{center}
{\bf Figure Captions}
\end{center}
\normalsize
\addtolength{\baselineskip}{\baselineskip}
\vskip 2 truecm
\par\noindent
{\bf Fig. 1:} Phase diagram of the pure folding problem ($L=\infty$).
The solid and the broken lines are respectively coexistence and critical lines.
\vskip 0.5 truecm
\par\noindent
{\bf Fig. 2:} The transition line separating at $h=0$ the disordered
folded phase ($M=0$) from the flat phases ($M=\pm 1$).
The solid and the broken lines are respectively first--order and critical
lines. The tricritical point is marked by a full circle.
\vskip 0.5 truecm
\par\noindent
{\bf Fig. 3:} Phase diagram of the folding model at $L=1.6$.
The solid and the broken lines are respectively coexistence and critical lines.
The full circles represent critical points (their coordinates are given in
Section 3). In the inset it is reported the magnification of the region of the
phase diagram in the empty square.
\vskip 0.5 truecm
\par\noindent
{\bf Fig. 4:} Phase diagram of the folding model at $L=1.3$.
The solid and the broken lines are respectively coexistence and critical lines.
The full circles represent critical points; their coordinates are:
$K=-0.1382,\; h=0.66$; $K=-0.1382,\; h=-0.66$; $K=0.31377,\; h=0$.
\vskip 0.5 truecm
\par\noindent
{\bf Fig. 5:} Phase diagram of the pure Ising model on a honeycomb lattice
($L=0$). The solid and the broken lines are respectively coexistence and
critical lines. The full circle represents the usual ferromagnetic Ising
critical point.

\newpage
\thispagestyle{empty}
\Large
\begin{center}
{\bf Table Caption}
\end{center}
\normalsize
\addtolength{\baselineskip}{\baselineskip}
\vskip 2 truecm
\par\noindent
{\bf Table 1:} Independent hexagon configurations.
\par\noindent

\newpage
\thispagestyle{empty}
\begin{center}
{\bf Table 1}
\end{center}
\medskip
\begin{center}
\begin{tabular}{|c|cccccc|c|}
\hline
Configuration & $s_1$ & $s_2$ & $s_3$ & $s_4$ & $s_5$ & $s_6$ &
Multiplicity \\
\hline
$z_1$ & $+$ & $+$ & $+$ & $+$ & $+$ & $+$ & 1 \\
$z_2$ & $-$ & $-$ & $-$ & $-$ & $-$ & $-$ & 1 \\
$z_3$ & $-$ & $+$ & $+$ & $+$ & $+$ & $+$ & 3 \\
$z_4$ & $+$ & $-$ & $+$ & $+$ & $+$ & $+$ & 3 \\
$z_5$ & $+$ & $-$ & $-$ & $-$ & $-$ & $-$ & 3 \\
$z_6$ & $-$ & $+$ & $-$ & $-$ & $-$ & $-$ & 3 \\
$z_7$ & $-$ & $-$ & $+$ & $+$ & $+$ & $+$ & 6 \\
$z_8$ & $+$ & $+$ & $-$ & $-$ & $-$ & $-$ & 6 \\
$z_9$ & $-$ & $+$ & $-$ & $+$ & $+$ & $+$ & 3 \\
$z_{10}$ & $+$ & $-$ & $+$ & $-$ & $+$ & $+$ & 3 \\
$z_{11}$ & $+$ & $-$ & $+$ & $-$ & $-$ & $-$ & 3 \\
$z_{12}$ & $-$ & $+$ & $-$ & $+$ & $-$ & $-$ & 3 \\
$z_{13}$ & $-$ & $+$ & $+$ & $-$ & $+$ & $+$ & 3 \\
$z_{14}$ & $+$ & $-$ & $-$ & $+$ & $-$ & $-$ & 3 \\
$z_{15}$ & $-$ & $-$ & $-$ & $+$ & $+$ & $+$ & 3 \\
$z_{16}$ & $+$ & $+$ & $+$ & $-$ & $-$ & $-$ & 3 \\
$z_{17}$ & $-$ & $-$ & $+$ & $-$ & $+$ & $+$ & 6 \\
$z_{18}$ & $+$ & $+$ & $-$ & $+$ & $-$ & $-$ & 6 \\
$z_{19}$ & $-$ & $+$ & $-$ & $+$ & $-$ & $+$ & 1 \\
$z_{20}$ & $+$ & $-$ & $+$ & $-$ & $+$ & $-$ & 1 \\
\hline
\end{tabular}
\end{center}

\end{document}